# High-temperature Anomalous Hall Effect in Transition Metal Dichalcogenide-Ferromagnetic Insulator Heterostructure

*Sheung Mei Ng, Hui Chao Wang, Yu Kuai Liu, Hon Fai Wong, Hei Man Yau, Chun Hung Suen, Ze Han Wu, Chi Wah Leung and Ji Yan Dai\**

Department of Applied Physics, The Hong Kong Polytechnic University, Hong Kong, 999077, P. R. China

E-mail: jiyan.dai@polyu.edu.hk

Integration of transition metal dichalcogenides (TMDs) on ferromagnetic materials (FM) may yield fascinating physics and promise for electronics and spintronic applications. In this work, high-temperature anomalous Hall effect (AHE) in the TMD ZrTe$_2$ thin film using heterostructure approach by depositing it on ferrimagnetic insulator YIG (Y$_3$Fe$_5$O$_{12}$, yttrium iron garnet) is demonstrated. In this heterostructure, significant anomalous Hall effect can be observed at temperatures up to at least 400 K, which is a record high temperature for the observation of AHE in TMDs, and the large R$_{AHE}$ is more than one order of magnitude larger than those previously reported value in topological insulators or TMDs based heterostructures. The magnetization of interfacial reaction-induced ZrO$_2$ between YIG and ZrTe$_2$ is believed to play a crucial role for the induced high-temperature anomalous Hall effect in the ZrTe$_2$. These results reveal a promising system for the room-temperature spintronic device applications, and it may also open a new avenue toward introducing magnetism to TMDs and exploring the quantum AHE at higher temperatures considering the prediction of nontrivial topology in ZrTe$_2$.

**Introduction**

Transition metal dichalcogenides (TMDs) are a family of two-dimensional (2D) layered materials like graphene. The TMDs have aroused enormous interests and studies of exploring their properties and applications. [1] Excellent and extraordinary properties were revealed especially when TMDs were exfoliated to atomically thin layers, [2] and this opened up opportunities for developing nanoscale devices in optical and electrical applications. [3] However, the nonmagnetic nature of most TMDs hinds the application of TMDs in spintronics, information memories and storage devices. Thus, introducing magnetic behavior in nonmagnetic transition metal dichalcogenides is essential to broaden their applications.

Extensive studies have been performed to investigate the feasible ways to induce and manipulate the magnetism in TMDs, [4] such as through defects, morphology fabricating, transition metal doping, external strain, ion intercalation and the van der Waals heterostructures. Among these approaches, the heterostructure method by placing the TMD material in close proximity to a ferromagnetic insulator (FMI), without introducing unnecessary disorder to the material or sacrificing its excellent intrinsic properties, is of significant interest. Such heterostructure approach allows direct manipulation and probe of magnetism via electrical method. Anomalous Hall effect is a momentous transport phenomenon in magnetic materials, [5] in which a transverse Hall voltage appears perpendicular to the applied current and the magnetization. Contrary to the well understanding of ordinary Hall effect due to Lorentz force, the underlying principle of the AHE has not been understood well though the related research has been conducted for more than one century. Intrinsic mechanism arising from the Berry curvature and extrinsic mechanism including skew scattering and side jump have been presented to explain the AHE although some debates needs further investigations. [5] As a typically observed feature in magnetic materials, AHE has been one of the important tools to detect and characterize the transport properties of the mobile electrons in ferromagnetic metals and semiconductors, and the effect shows great potential in applications especially in the spintronics.

Zirconium di-telluride (ZrTe$_2$) is a non-magnetic TMDs as well as a candidate of topological materials showing excellent properties promising for quantum devices applications, such as high mobility and large magnetoresistance. [6, 7] Thus, introducing magnetism in the TMD ZrTe$_2$ by heterostructure approach proximity to a ferromagnetic insulator is rather interesting since the quantum AHE (QAHE) is predicted in the integrated systems of topological materials and ferromagnetic materials. [8] The quantum AHE possesses dissipation less chiral edge current transport in the absence of external magnetic field, which promises spintronic applications at the nanoscale. [9] To date, one of the most important topics in this field is the realization of high temperature QAHE, [10] for which the magnetic behavior persisting to high temperatures is an essential prerequisite. In this work, we demonstrate induced magnetism in ZrTe$_2$ via the heterostructure approach and directly probe the magnetism by measuring the AHE. A large AHE was observed in wide-range temperatures (10-400 K) for the 4-nm thick ZrTe$_2$ film. Our work presents a promising candidate for the room temperature applications in spintronics. Especially, the temperature enhanced results may pave the way for realizing QAHE-based devices considering the predicted topological category of ZrTe$_2$.

**(Main text)**

**Figure S1**a shows XRD pattern of the grown YIG film, where one can see that the YIG film is epitaxially grown on GGG substrate presenting only the (110) peak. Based the XRD result, the lattice constants of the YIG film and GGG substrate are calculated to be 12.45Å and 12.37Å respectively, indicating a small lattice mismatch of ~0.6% between them. The ω-scan (Figure S1b) shows the full-width at half-maximum (FWHM) of 0.07°, suggesting excellent crystallinity of the YIG film. The AFM image shown Figure S1c presents very clear terraces,

indicating a clean and atomically flat surface. Figure S1d shows the room-temperature magnetization measurement results of the YIG/GGG with the magnetic field oriented in both in-plane and out-of-plane directions. One can see the well-defined in-plane magnetic anisotropy, and the hysteresis loops give the in-plane and out-of-plane coercive fields at ~3 Oe and ~100 Oe, respectively.

**Figure 1** shows cross-sectional TEM images of the heterostructures composing different $ZrTe_2$ thicknesses ranging from 4 to 30 nm. It is apparent that the YIG layer is parallel to its (110) atomic plane and the layered structure of $ZrTe_2$ films are formed in all samples with columnar grains and preferred [0001] growth orientation. An amorphous layer can also be observed next to the crystallized YIG layer, and in all samples, the total thickness of the crystallized YIG layer and the amorphous layer matches the thickness of deposited YIG layer. From the energy dispersive X-ray (EDX) spectrum (**Figure S2**), this layer contains Te beyond the composition Y and Fe, suggesting the possibility of diffusion of Te into YIG during $ZrTe_2$ deposition. Therefore, this amorphous layer is attributed to Te doped YIG (a-YIG:Te).

It is also apparent that, before forming the layer-structured $ZrTe_2$, there is a transition layer formed between the a-YIG:Te and the crystalized $ZrTe_2$. According to the lattice parameters and composition obtained (see Figure. S2), this transition layer is attributed to be $ZrO_2$ resulted from the out diffusion of oxygen from the YIG layer to react with Zr during early stage of $ZrTe_2$ growth. The $ZrO_2$ layer is a high-$k$ insulator and it does not contribute to any electrical signals. During the last decade, both experimental and theoretical studies have shown that intrinsic defects in the oxide nanomaterials can lead to room temperature ferromagnetism.[11, 12] A recent paper has reported that oxygen vacancy defect-rich $ZrO_2$ nanostructures show high $T_c$ (700 K) and high magnetization (5.9 emu/g).[12] It should be noted that the deposition process of $ZrTe_2$ takes place at high vacuum environment, and therefore, the $ZrO_2$ layer formed under such a process should be rich in oxygen vacancies which are possibly responsible for the magnetic moments. Thus, the unexpected polycrystalline $ZrO_2$ instead of the intended YIG can actually play the role of ferromagnetic insulator in the heterostructures, which may introduce magnetic order to the TMD $ZrTe_2$ by interfacial exchange coupling due to the direct contact.

**Figure 2**a is a schematic diagram of the sample configuration for electrical transport measurements. Temperature-dependent sheet resistance ($R_{sheet}$-$T$) of the heterostructures are shown in Figure 2b, where one can see that the $ZrTe_2$ thin films ranging from 4-30 nm show semiconducting behavior. Our earlier results show that the 60 nm-thick $ZrTe_2$ thin films on single crystalline STO are metallic.[7] As we know, the thickness can be an important parameter to tune the electronic properties of TMDs. The semiconducting feature may arise from an opened gap in the TMD due to the quantum confinement when the thickness is decreased. However, the first-principles calculations of the band structure by density functional theory reveal semimetal feature of $ZrTe_2$ monolayer, which excludes the above possibility. On the other hand, the metal-semiconductor transition may be due to the increase in disorder with decreasing layer number.[13] We also performed the transport measurements of a 10 nm-thick $ZrTe_2$ thin film on $SiO_2$/Si as shown in Figure 2c, which shows similar semiconducting behavior as the heterostructure. It is suggested that the semiconducting behavior in the very thin $ZrTe_2$ films within the heterostructures is due to the increased localization induced by the disorder, defects and even oxygen intercalations when grow on the some relatively volatile oxide substrates.

**Figure 3**a illustrates the Hall measurement results at low temperatures from 2-50 K for the heterostructure consisting of 18 nm-thick $ZrTe_2$ thin film. Obvious squared Hall hysteresis loops are observed at 2 K, 5 K and 10 K. At 50 K, the hysteresis loop is too small to be detected because of the insufficient signal-to-noise ratio. This type of loop in a normal ferro- or ferrimagnetic conductor is the manifestation of typical AHE identifying the existed magnetism. In the constructed heterostructure, both the ferromagnetic YIG and $ZrO_2$ layers are insulators, and thus the Hall response must come from the $ZrTe_2$ thin film. The corresponding magnetoresistance (MR) curves of the heterostructure at selected temperatures are further shown in Figure 3b. One can see that the MR is negative with a butterfly shape hysteresis loop. The hysteresis loop shows that the two separate maxima of MR are most prominent at 2 K with $H_{max} = \pm \sim 3500$ Oe. As the temperature is increased above 50 K, the loop is gradually obscured by the background MR signal and cannot be clearly resolved. The behaviors of the heterostructures are in sharp contrast to that of the $ZrTe_2$ thin film on $SiO_2$/Si which shows linear Hall resistance and non-hysteretic positive MR (Figure 3c and 3d). The magnetic features in the Hall and MR results of the heterostructures indicate the presence of magnetic order in the TMD film related to the underlying ferromagnetic materials. The negative MR was suggested to arise from the weak localization effect with magnetic scattering around the interface when adjacent to a ferromagnetic insulator.[14] Thus, the nonlinear Hall should predominantly originate from the anomalous Hall effect as a result of proximity effect.[15]

In the AHE framework, the overall Hall resistance ($R_{yx}$) (**Figure S3** a) is a combination of the ordinary Hall effect (OHE) and the anomalous Hall effect:[5] $R_{yx} = R_H(B) + R_{AHE}(M)$. The ordinary Hall effect $R_H(B)$ can be removed by subtracting the linear background (see supplementary Figure S3). It is found that the AHE in the 18 nm-thick $ZrTe_2$ film becomes very weak when the temperature is increased to room temperature (**Figure S4**). However, when the film thickness is decreased, the AHE at 300 K can be enhanced. **Figure 4**a shows the thickness dependence of the $R_{AHE}$ presented by the heterostructures. One can see that the room temperature $R_{AHE}$ vs. H plot shows the S-shaped curves with decreasing trend upon increasing film thickness. To clearly compare the effect of thickness, $R_{AHE}$ at 50 kOe for each thickness were extracted as shown in Figure 4b. It is apparent that the $R_{AHE}$ decreases dramatically with increased thickness for the curves measured at both room temperature and low temperature. With $ZrTe_2$ thickness increased from 4 to 30 nm, the room-temperature $R_{AHE}$ decreases from 24.4 to 0.004 Ω while $R_{AHE}$ at 100 K decreases from 33.7 to 0.008 Ω. It should also be noted that for all thicknesses, $R_{AHE}$ shows temperature dependent nature that $R_{AHE}$ is further enhanced at lower temperature and the curves show saturation at comparable magnetic field strength. The thickness dependence of the AHE suggests that mainly the interface is modulated by the underlying ferromagnetic materials.

Figure 4c shows the anomalous Hall resistance of the heterostructure consisting of 4 nm-thick $ZrTe_2$ thin film. Significant AHE was observed in wide-range temperatures from 10 to 400 K. The $R_{AHE}$ is large with a value of ~40 Ω at 10 K and ~20 Ω at 400 K. Figure 4d shows the saturation values of $R_{AHE}$ vs. temperature, indicating that the AHE can be observed at even above-400 K. To our knowledge, this is a record high temperature for the observation of AHE in TMDs, and the large $R_{AHE}$ is more than one order of magnitude larger than those previously reported value in topological insulators or TMDs based heterostructures [10]. The deposition of $ZrTe_2$ on YIG yields unexpected extra layers and complicated heterostructure of $ZrTe_2/ZrO_2$/a-YIG:Te/YIG. It can be expected that the ferromagnetic $ZrO_2$ formed in the heterostructures plays an important role for the induced magnetic behavior in the $ZrTe_2$ since it is in directly contact to the $ZrO_2$ thin film. In particular, the existence of high temperature AHE in the heterostructure may be attributed to the high $T_c$ of $ZrO_2$.[12]

**Conclusion**

In summary, our work demonstrates AHE in the TMD $ZrTe_2$ thin films beyond room temperature by using a heterostructure approach. Large AHE has been observed in wide temperatures from 10 to 400 K for the heterostructure consisting of 4 nm thick $ZrTe_2$ film. As the thickness of $ZrTe_2$ increases, the AHE decreases sharply. We attribute these transport results to the interfacial exchange coupling between the TMD ZrTe$_2$ and the ferromagnetic ZrO$_2$. The impressive high temperature for the AHE paves way for realizing the quantum AHE at higher temperatures considering the possible topological nature of ZrTe$_2$. The work provides a system with promises for innovative electronic and spintronic applications at room temperature and beyond. These findings due to an unexpected complex interface in the heterostructure system will be both enlightening in further research studies and prospective in technological applications of integrated devices. Further investigations such as X-ray magnetic circular dichroism would be helpful to construct a better understanding on the origin of AHE.

**Experimental Section**

The ZrTe$_2$/YIG heterostructure in the work was prepared by pulsed-laser deposition technique with a KrF excimer laser (emission wavelength = 248 nm). Firstly, about 30 nm-thick YIG film was grown on atomically flat (110) GGG (Gd$_3$Ga$_5$O$_{12}$, gadolinium gallium garnet) substrate at substrate temperature of 710 °C with laser energy ~200 mJ and oxygen pressure of 13 Pa. After deposition, *in-situ* annealing was carried out at O$_2$-rich environment for 10 mins. The YIG is a ferrimagnetic insulator with a high Curie temperature (T$_c$~550 K), so that it does not cause any current shunting in the electrical measurements. After that a layer of ZrTe$_2$ of various thicknesses was deposited onto the YIG/GGG at high vacuum (10$^{-4}$ Pa) with substrate temperature of 400 °C and laser energy 250 mJ. The film was grown into a Hall bar structure (width: 165 μm, length: 390 μm) by deposition through a stainless steel shadow mask. Finally, a layer of AlN capping was deposited on top of the Hall bar structure. Structural characterizations were performed with X-ray diffractometer (XRD, Rigaku SmartLab) and transmission electron microscope (TEM, Jeol JEM 2100F). Morphological study was conducted with atomic force microscope (AFM, Bruker NanoScope 8). Magnetic and electrical transport properties of the samples were investigated with a vibrating sample magnetometer (VSM, Lakeshore 7410) and physical property measurement system (PPMS, Quantum Design Model 6000).

**Supporting Information**
Supporting Information is available from the Wiley Online Library or from the author.

**Acknowledgements**

This research was supported by Hong Kong GRF grant (153094/16P). Financial support from The Hong Kong Polytechnic University strategic plan (No: 1-ZE25 and 1-ZVCG) and postdoc fellow scheme (1-YW0T). Research Grants Council, HKSAR (PolyU 153027/17P) was also acknowledged. Sheung Mei Ng and Hui Chao Wang contributed equally to this work.

**Figure 1.** Cross-sectional TEM imaging of the $ZrTe_2$/YIG heterostructure of different $ZrTe_2$ thicknesses, ranging from (a) 4nm, (b) 9nm, (c) 18nm and (d) 30nm. Planes of $ZrTe_2$ (0001) were indicated in the crystallized layer and the interfaces between each layer were indicated with the arrows. Inset in (b) highlights one unit cell of $ZrO_2$.

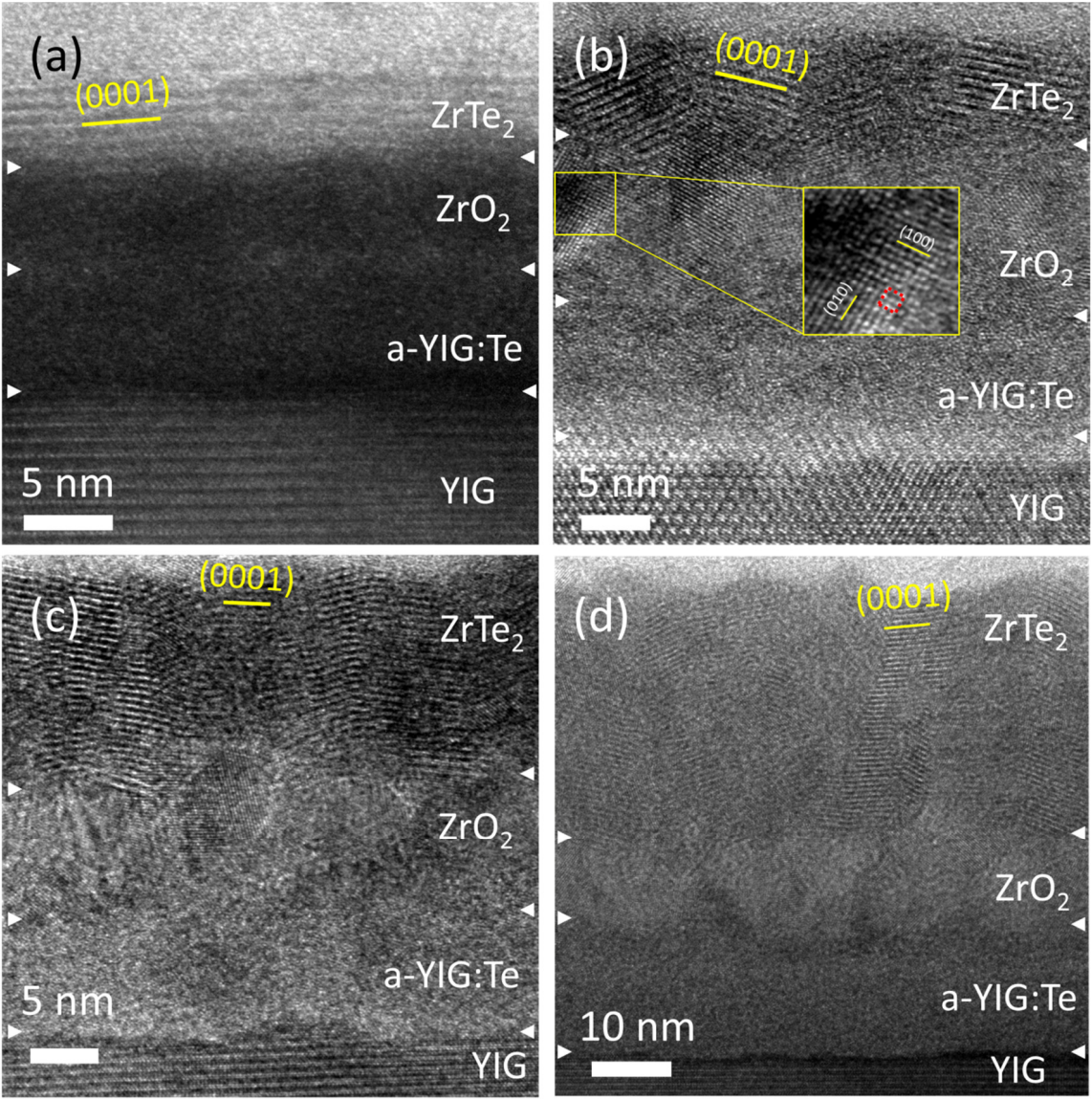

**Figure 2.** (a) a schematic diagram for measurement. Temperature dependence of sheet resistance of ZrTe₂ grown on YIG (b) and SiO₂/Si (c) respectively.

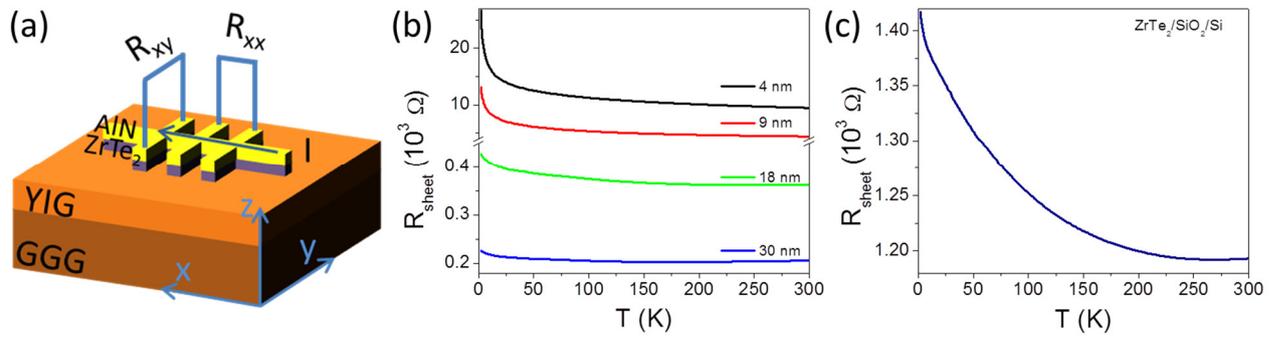

**Figure 3**. (a) anomalous Hall effect and (b) magnetoresistance in 18-nm thick $ZrTe_2$ within the heterostructure. (c) $R_{yx}$ of 10-nm thick $ZrTe_2/SiO_2/Si$ measured at different temperatures showing linear ordinary Hall. (d) Magnetoresistance of $ZrTe_2/SiO_2/Si$ at 10K. Curves in (a)-(c) are shifted for clarity.

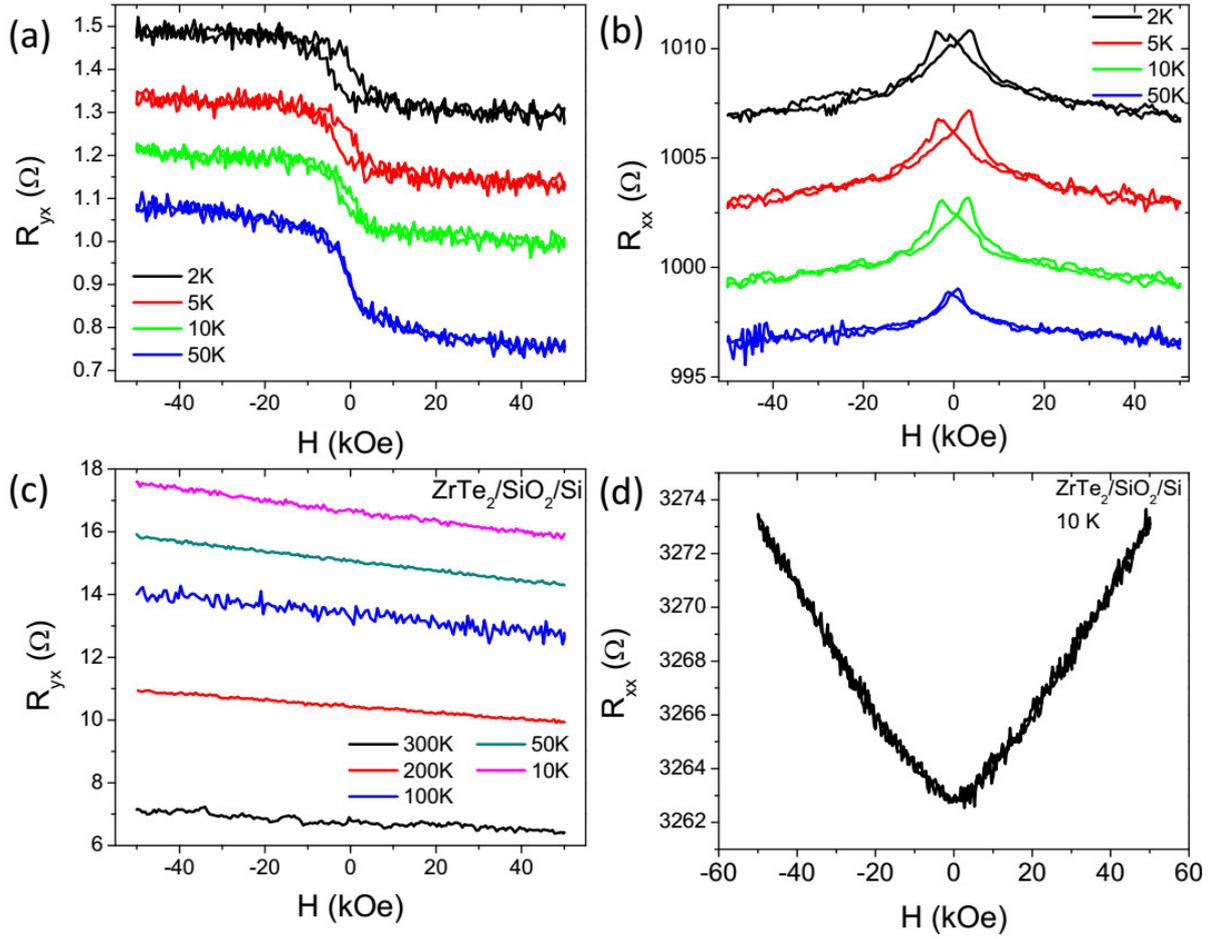

**Figure 4**. (a) Anomalous Hall effect of the heterostructures consisting ZrTe$_2$ thin film with different thickness at 300 K. (b) Thickness dependence of the extracted $R_{AHE}$ at high magnetic field. (c) $R_{AHE}$ of the 4-nm ZrTe$_2$ within the heterostructure at 300K. (d) $R_{AHE}$ as a function of measurement temperatures from 400K to 10K.

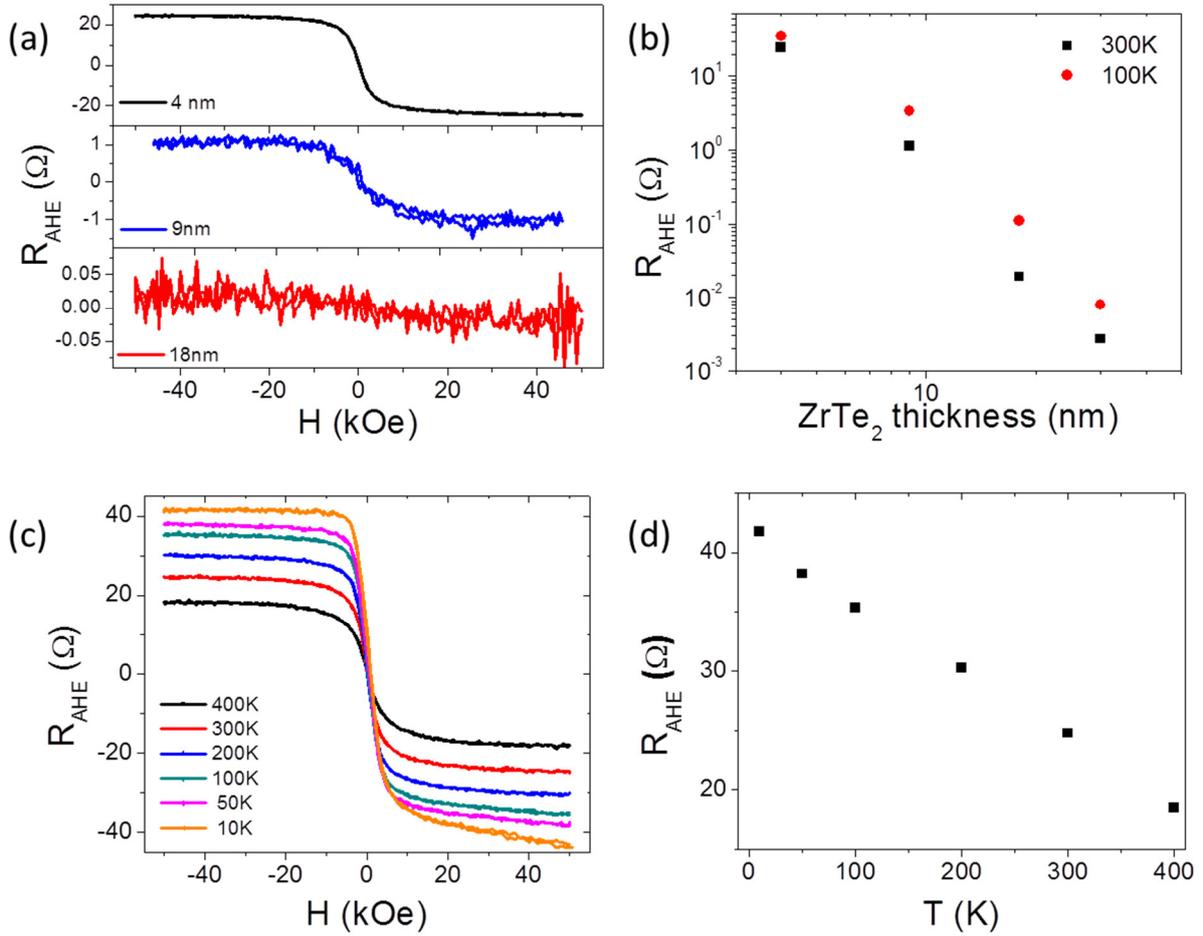

# Supporting Information

**High-temperature Anomalous Hall Effect in Transition Metal Dichalcogenide-Ferromagnetic Insulator heterostructure**

*Sheung Mei Ng, Hui Chao Wang, Yu Kuai Liu, Hon Fai Wong, Hei Man Yau, Chun Hung Suen, Ze Han Wu, Chi Wah Leung and Ji Yan Dai\**

Department of Applied Physics, The Hong Kong Polytechnic University, Hong Kong, 999077, P. R. China

E-mail: jiyan.dai@polyu.edu.hk

**Figure S1.** (a) 2θ/ω scan and (b) ω –scan of YIG/GGG. (c) AFM image of YIG/GGG (z = 5nm). (d) Normalized magnetic hysteresis loops of a YIG/GGG film at 300 K with in-plane and out-of-plane fields.

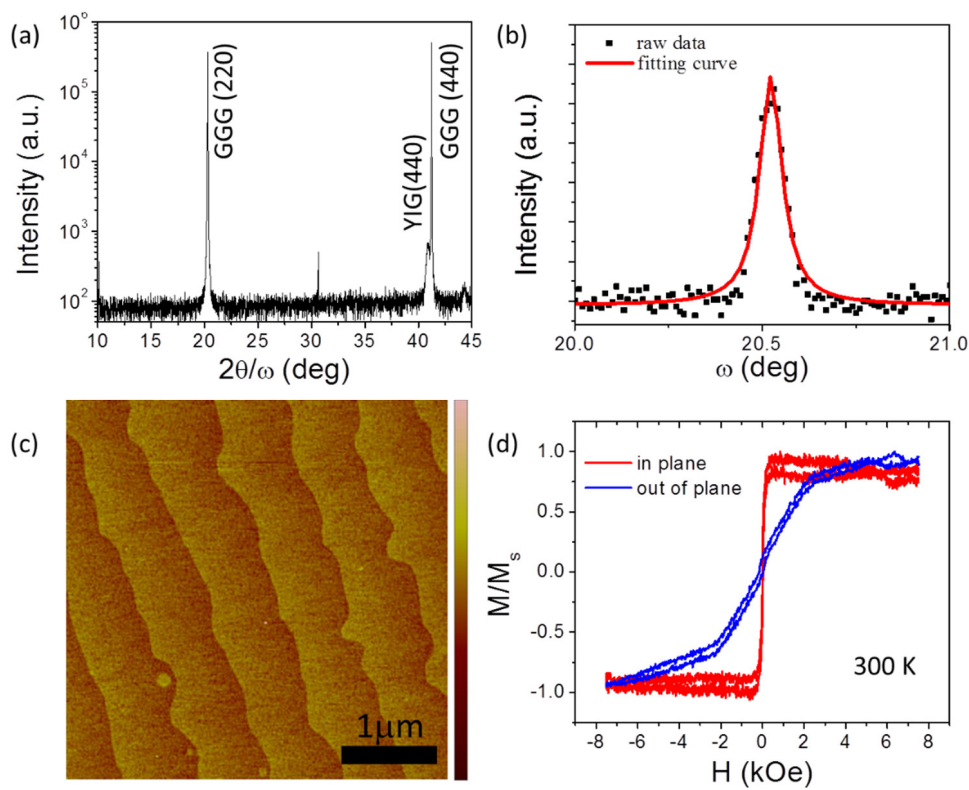

**Figure S2**. EDX spectrum was obtained from the region of amorphous layer. The presence of Y and Fe suggests the composition related to YIG (O is not shown) while the presence of Te suggests the diffusion from ZrTe$_2$ (Gd is from substrate GGG). The EDX result suggests the layer is amorphous YIG with Te.

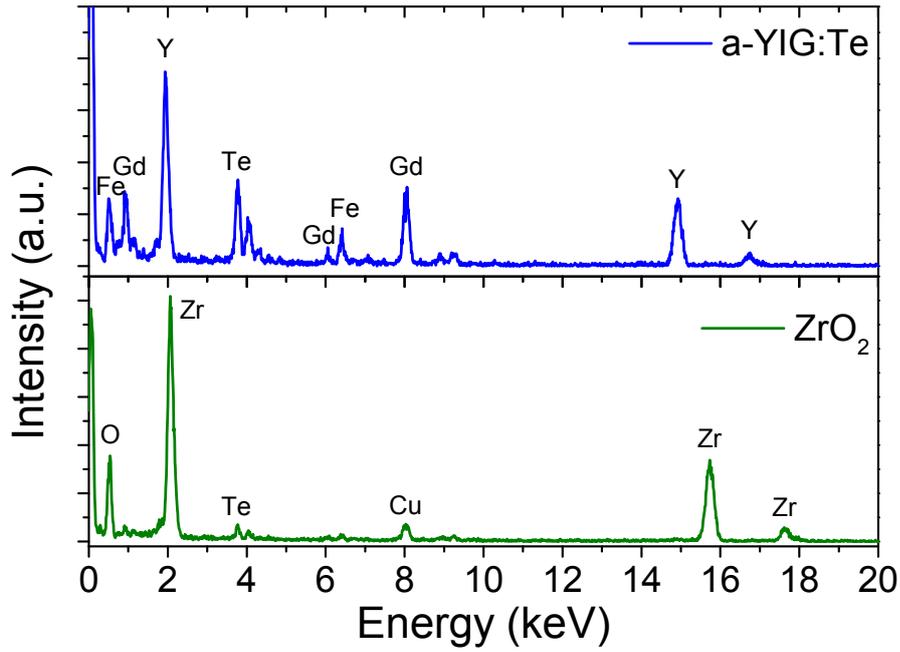

**Figure S3.** The anomalous Hall effect ($R_{AHE}$) was obtained by subtracting the linear background, which is contribution of ordinary Hall effect as indicated by the red dashed line in (a), from the total Hall resistance ($R_{yx}$).

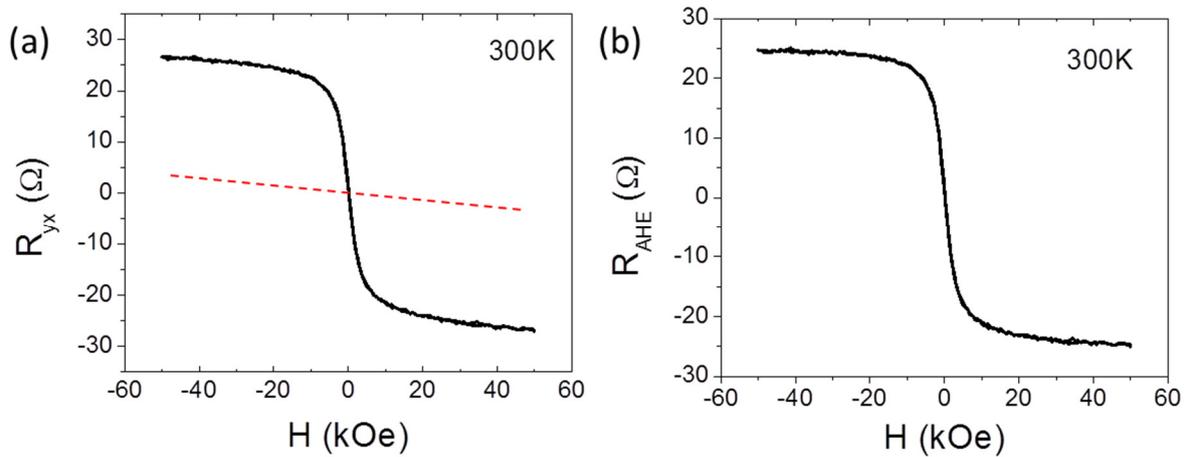

**Figure S4.** The anomalous Hall effect ($R_{AHE}$) at selected temperatures for the 18 nm thick $ZrTe_2$ thin film within the heterostructure. The curves are shifted for clarity.

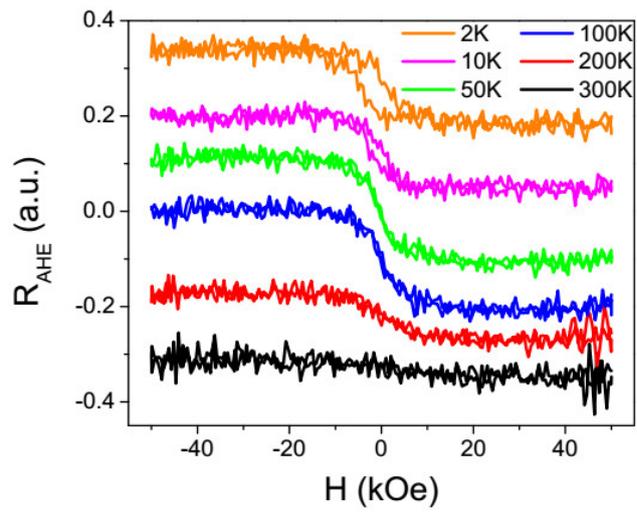